\documentclass[a4paper, twocolumn]{article}
\makeatletter
\def\pdfstartlink@attr{}
\makeatother
\usepackage{amsmath, amsthm, bm, mathrsfs, amscd, amssymb, graphicx}
\usepackage{enumerate} \usepackage{authblk}
\usepackage[hidelinks]{hyperref}
\hypersetup
{
    pdfauthor={A. Shariati, N. Jafari},
    pdfsubject={Alternative theories of gravity},
    pdftitle={Modified Inertia as non-conservative Newtonian dynamics},
    pdfkeywords={gravity, modified gravity, binary systems}
    pdfborder={0 0 0},
colorlinks=true,
 linkcolor=blue,
urlcolor= blue,
filecolor=blue,
citecolor={blue}, 
urlbordercolor = {0 0 0},
citebordercolor = {0 1 0},
linkbordercolor = {1 0 0},
  linktoc=all,
}
\usepackage[varg]{txfonts} 

\newcommand{\abs}[1]{\left\vert #1 \right\vert}

\providecommand{\apj}{ApJ}
\providecommand{\prl}{Phys. Rev. Lett.}
\usepackage{natbib}
\newcommand{\LagDensity}{\mathscr{L}}
\newcommand{\Order}[2]{{\cal O}\left(\frac{#1}{#2}\right)}

\newcommand{\BMF}{\mathscr{F}}
\newcommand{\reducedmass}{\mu}
\newcommand{\Fmond}{{\cal F}}
\newcommand{\abssmall}[1]{\vert #1 \vert}
\newcommand{\oder}[2]{\frac{{\rm d}#1}{{\rm d}#2}}
\newcommand{\pder}[2]{\frac{\partial #1}{\partial #2}}
\newcommand{\sun}{\odot}
\renewcommand{\Hat}[1]{\ensuremath{\hat{{\bm{#1}}}}}
\renewcommand{\vec}[1]{\ensuremath{{\bm{#1}}}}

\title{Modified Inertia as Nonconservative Newtonian Dynamics}
\author[1]{Ahmad Shariati\thanks{a.shariati@alzahra.ac.ir}}
\affil[1]{\small Department of Physics, Faculty of Physics and Chemistry, 
              Alzahra University, Tehran 19938 97973, Iran.}
\author[2]{Nosratollah Jafari\thanks{nosratollah.jafari@nu.edu.kz}}
\affil[2]{Department of Physics, Nazarbayev University, Kabanbay Batyr Ave 53, 
              Nur-Sultan, 010000,   Kazakhstan.}
\date{19 Apr 2021}

\begin{document}
\maketitle
\begin{abstract}
Modified Newtonian Dynamics of Milgrom (MOND) is a paradigm for explaining the rotation 
curves of spiral galaxies and various other large scale structures.  This paradigm includes several
different theories.   Here we present Milgrom's modified inertia (MI) theory in terms of a simple and tractable
non-conservative Newtonian dynamics, which is useful in obtaining observable predictions of MI.
It is found that: 1) Modified inertia theory is equivalent to a Newtonian theory, 
with a non-conservative gravitational field, and dark matter density. 2) The tidal force 
in the equivalent Newtonian dynamics is non-conservative,  and its effect on a binary 
system in free fall in the gravitational field of a spheroid is addressed. 
We also discuss attempts to restore conservations in MI.
\end{abstract}

\noindent Published 19 Oct 2021 in 
\href{https://journals.aps.org/prd/abstract/10.1103/PhysRevD.104.084070}{Phys. Rev D \textbf{104}, 084070}
\section{Introduction} 
To solve the plateau of rotation curves of spiral galaxies  there
are two paradigms, Dark Matter (DM), and Modified Newtonian
Dynamics (MOND) which was proposed by \citet{Milgrom-1983ApJ, Milgrom-1983b}
and has been vastly investigated by Milgrom himself 
\citep{Milgrom-1994a, Milgrom-1999, Milgrom-2002a, Milgrom-2009a, Milgrom-2010a, Milgrom-2012a, Milgrom-2014a}
and various other authors \citep{BM-1984, Bekenstein-2004}.
For a review of MOND and dark matter we refer the reader to 
\citep{Milgrom-2001a, SR2001, FM2012a, Milgrom-2020a, SM2002}.

MOND in not a single theory, it refers to several theories the basic idea behind them is if one can
change Newtonian dynamics on the scale of galaxies, so that the plateau of the rotation curves could 
be explained with no need to add dark matter halos.  
The original version of \citet{Milgrom-1983ApJ}  is the \emph{modification of inertia} (MI)
in the form $m\, \mu(a)\, \vec a = \vec h$,
where $\vec h = -\vec\nabla\phi$ is the Newtonian gravitational field.
Later,  \citet{BM-1984} studied the so called 
\emph{modified gravity} (MG) theories, which keep $m\, \vec a = \vec F$ intact, but
the gravitational field is not the Newtonian one.  There appeared several modified gravity theories
which try to implement Milgrom's idea.  For a comparison of these theories see \citet{Zhao2010},
where the authors use a generalized virial theorem to compare different modified gravity theories.


In this article, we would like to present a system, which is mathematically equivalent to Milgrom's MI.
The benefit of this equivalent system is to provide a framework for finding predictions of the 
MI---predictions that could be used for verification or refutation of the theory.
This is just a first step towards simplifying MI, just like
quasi linear MOND made MOND easier \citep{Zhao2010, ZLB-2010}, 
with the possible advantage of tackling the non-conservativeness challenge of MI mathematically easier. 

The original idea of modifying Newtonian dynamics is to write the governing differential
equations not as the usual Newtonian form $\vec{F} = m \,
\vec{a}$, but as the equation $\vec{\Fmond} = m \, \vec{a} \, \mu(a)$,
where $\mu$ is a function characterizing the theory (to be discussed bellow).
It should be noted that we are deliberately using $\Fmond$ instead of
$F$:  While $\vec\Fmond$ denotes the force field in the MOND framework,  $\vec F$ 
denotes the force filed in the Newtonian framework.

The function $\mu(a)$, called the interpolating function, is a monotonically increasing function 
of the absolute value of the acceleration, $a = \abssmall{\vec{a}}$, such that for large enough 
values of $a$ (compared to some fundamental acceleration of the theory, denoted by $a_0$), 
$\mu(a) \simeq 1$, and for very small values of $a$, $\mu(a) \simeq 0$. 
Milgrom  showed that this modification of the
Newtonian dynamics, for $a_0 \sim 10^{-10}\, {\rm m}\,{\rm s}^{-2}$
could account for the flatness of the rotation curves of spiral
galaxies, with no need of introducing any extra \emph{dark} matter.

To find which one of these paradigms is chosen by the nature, various
groups proposed or did experiments \citep{ZLB-2010, Ignatiev-2015}. 
 \citet{GSSC2007} have shown that, for accelerations as small as
$10^{-14}\, {\rm m}\,{\rm s}^{-2}$ the Newtonian equation $\vec{F} =
m\, \vec{a}$ is valid.  
Existance of galaxies without dark matter is another reason against MOND 
\citep{Dokkum:2018td}, though there are arguments against such reasoning
\citep{Kroupa:2018vi}, it seems reasonable that the rotation curves support standard matter
\citep{KBH2018}.  
\citet{McGaugh-2016} reported a correlation between the radial acceleration traced by
rotation curves and the observed distribution of baryonic matter, which might be construed as
a signal against dark matter paradigm (since this means dark matter must somehow be correlated to
baryonic matter).  But we should note that MOND is not the only way to interpret this correlation.
In another direction, recently \citet{Chae-2020} published evidence for the violation of the strong equivalence principle in favor of MOND.

To interpret observations and to predict observable phenomena, 
we have to have theoretical framework.  In this article,
we emphasize on a theoretical argument, based on a transformation from MOND differential 
equation of a test particle to  Newtonian equations, which leads to various conclusions and
predictions in the framework of MOND.
We use a transformation to write the modified inertia (MI) equation $\vec h = \mu(a)\, \vec a$ as
a Newtonian equivalent form $\vec a = \vec g$, where $\vec g$ is obtained from $\vec h$ unambiguously.
We then use this equivalent version to study the consequences of modifying inertia, and
to study the relation between MI and MG
versions.  We found that modifying inertia leads to non-conservation of energy in
a binary system which is in free fall in the halo of the galaxy.  This result could not be
derived by the method of \citet{Zhao2010}, the results of which are valid for the 
modified gravity theories.

\section{Transforming to an Equivalent Newtonian Theory} \label{sec:dual}
The basic idea behind this article, is that $\vec{F} = m\, \vec{a}$ \emph{is a framework to write
the dynamics} \citep{Wilczek2004a}.
Newtonian dynamics (ND) is based on the differential equation
\begin{equation}  \label{ND}
  \text{ND} \qquad  \vec{F} = m\, \vec a,
\end{equation} 
where
\begin{math} \vec a = {\rm d}^2\vec r/{\rm d}t^2 \end{math} is the acceleration.
The most important feature of this  is that the differential equations governing a
point particle are of second order, such that when put in the form $\vec{F} = m\, \vec{a}$, the 
\textit{Newtonian force}
$\vec F$ depends on the position and the velocity of the particle, and not on the 
acceleration $a$ itself.  In classical electrodynamics, the reaction force of a radiating 
charged particle violates this assumption, but that's beyond our present considerations.
In a gravitational field $\vec g$, the motion of a test particle is given by the equation 
$ \vec a = \vec g$,  where
where $\vec g$ is determined by the  field equations,
$\vec\nabla\times\vec g  = 0$ and  $ \vec\nabla\cdot\vec g  = -4\, \pi\, G\, \rho$,
 $\rho$ being the mass density of the source.

 MI version of MOND states that the differential equations for a test particle of 
 mass $m$ are
\begin{equation} \label{MOND}
  \text{MOND} \qquad \vec{\Fmond} = m \, \vec{a} \, \mu(a),
\end{equation}  
where the \emph{interpolating} function $\mu(a)$ is a dimension-less, smooth, positive, 
and monotonically increasing function of the absolute value of the acceleration 
$a = \abssmall{\vec a}$, depending on a fundamental small acceleration 
$a_0 \sim 10^{-10}\, {\rm m}\, {\rm s}^{-2}$, having the following properties:
\begin{align} \label{eq:mu-function1}
\lim_{a\to 0^+} \mu(a) & = 0, \\  \label{eq:mu-function2}
 \lim_{a\to\infty} \mu(a) & = 1, \\ \label{eq:mu-function3}
  \mu'(a) & > 0 \quad \forall a > 0.
\end{align}
As \citet{Milgrom-1983ApJ} stated  explicitly: 
``The force field $\vec\Fmond$ is assumed to depend on its sources and to 
  couple to the body, in the conventional way.'' 
Consider the MI equation for a gravitational field $\vec{h}$,
where $\mu(a)$ is the interpolating function which defines the MI,
and $\vec{h}$ solves the usual Newtonian field equations. Thus, the MI version of MOND is
given by the following system:
\begin{align} 
\label{eq:mond1}   \vec a\, \mu(a) & = \vec h, \\
\label{eq:mond2}   \vec\nabla\cdot\vec{h} & = - 4 \, \pi \, G \, \rho_{\rm m}, 
\\ \label{eq:mond3}   \vec\nabla\times\vec{h} & = 0, 
  \end{align}
Here $\rho_{\rm m}$ is the \textit{mass density} function in the MOND framework.

It is easy to see that  \begin{math}\vec\Fmond = m\, \vec a\, \mu(a)\end{math}
 can  be transformed to  \begin{math} \vec F = m\, \vec a\end{math}.
 To do this, we first introduce the \emph{pseudo-acceleration} field
\begin{equation} \vec h = m^{-1}\, \vec\Fmond, \end{equation} 
and write MOND equations thus:
\begin{equation} \vec h = \vec a\, \mu(a). \end{equation}
Since $\vec h$ and $\vec a$ are parallel,  we have
\begin{math} h = a\, \mu(a)\end{math}.
Using the inverted interpolating function $\nu(h)$ (see appendix \ref{appendix1}),
we write this as $a = h\, \nu(h)$.  Multiplying it with $\Hat{a} = \Hat{h}$ (the unit vector)
we get
\begin{equation}\vec{a} = \vec{h} \, \nu(h).\end{equation} 
We can also multiply by $m$ to get
\begin{equation} m\, \vec{a} = \vec\Fmond\, \nu\left(\tfrac{\Fmond}{m}\right) = \vec F. \end{equation}
This is the usual
Newton's equation of motion, for \emph{the acceleration field}
\begin{equation} \label{eq:gdef} \vec g = \vec{h} \, \nu(h). \end{equation}
Thus, the modified inertia equation 
$m\, \vec{a} \,\mu(a) = \vec{\Fmond}$ could be written as $m\,\vec{a} = m\,\vec g$, 
where  $\vec h = m^{-1}\, \vec\Fmond$ is what we call the \emph{pseudo-acceleration field},
and 
$\vec{g} := \vec{h} \, \nu(h)$ is
\emph{the  acceleration  field}.

Since we have good  experience with the
Newtonian equation $\vec a = \vec g$, and because it is completely equivalent to the
modified inertia equation $\vec{a} \, \mu(a) = \vec{h}$, we could now derive
some useful information about modified inertia (MI) theories.

Defining $\vec g$ by (\ref{eq:gdef})
 the MI version of MOND equation 
of motion of a test particle is equivalent to the Newtonian one 
\begin{align}
 \label{eq:demond1} \vec a & = \vec g.
\end{align}   
Using (\ref{eq:mond2}, \ref{eq:mond3}, \ref{eq:gdef}) it is easy to find field equations governing 
$\vec{g}$:
\begin{align} 
\label{eq:demond2} \vec\nabla\cdot\vec{g}  &=  - 4 \, \pi \, G \, \rho_{\rm m} \, \nu(h) +
\vec{h} \cdot \vec\nabla\nu(h),
\\ \label{eq:demond3} \vec\nabla\times\vec{g} & =  - \vec{h} \times \vec\nabla \nu(h).
\end{align}
So the dynamics of a test particle in MI version of MOND is equivalent to the system 
of equations (\ref{eq:demond1}-\ref{eq:demond3}), where $\vec h$ is the solution to
(\ref{eq:mond2}, \ref{eq:mond3}).

\section{Physical Implications}
We are now going to obtain information about the MI version of MOND, using system
(\ref{eq:mond2}, \ref{eq:mond3}, \ref{eq:demond1}-\ref{eq:demond3}).
\subsection{Dark matter in disguise.} 
We see that in the Newtonian dynamics equivalent version,
 the mass density is being modified (multiplied by
$\nu(h)$), and we have got an extra term in the right-hand-side of
$\vec\nabla\cdot\vec{g}$, which could be interpreted as a
\textit{dark} mass density.   
\begin{align} \rho_{\rm d} & = -\frac{1}{4\,\pi\, G} \vec h \cdot \vec\nabla \nu(h)
 = -\frac{1}{8\,\pi\, G}\cdot\frac{\nu'(h)}{h}\, \vec h \cdot\vec\nabla h^2. 
\end{align}
For the simple form of the function $\mu(a)$, from (\ref{eq:etaprime-simple}) we get
\begin{equation} \label{eq:rhod}
\rho_{\rm d} = \frac{1}{8\,\pi\, G}\cdot\frac{a_0}{h^2\, \sqrt{h^2 + 4\, a_0\, h}} \, \vec h \cdot 
\vec\nabla h^2.
\end{equation}
By assuming that the acceleration due to visible and dark matter are always parallel,
\citet{Dunkel2004}  showed that the MOND equations 
can be derived from classical Newtonian dynamics, provided one also takes into account the gravitational 
influence of a DM component. \citet{Sivaram2017} took a similar approach.
The approach of the present article however is more general and shows that dark matter
is an  an inevitable consequence of modifying inertia.  This will be more clear in the
following section.

\subsection{Dynamics around a point particle.}
As an example, let's consider the
acceleration around a point mass $M$. Here
\begin{math}\vec{h} = -(G\, M/r^2)\, \Hat{r} \end{math}, for which 
\begin{math} h = {G\, M}/{r^2} \end{math}, and 
using the simple form of the MOND function $\mu$
 we get 
\begin{align}
 \vec\nabla\cdot\vec{g} 
& = -4\,\pi\,G\, M \, \delta(\vec{r}) - \frac{2\, a_0}{r}
 \, \left(1 + \frac{4\, a_0\, r^2}{G\, M}\right)^{-1/2}, \\ 
\vec\nabla\times\vec g & = 0.
\end{align}
This clearly shows that accepting modified inertia
equation $\vec{a} \, \mu(a) = \vec{h}$ for a point particle,
is equivalent to accepting an infinite dark matter, with density 
\begin{equation} \label{pp-dmA}
\rho_{\rm d}= \frac{a_0}{2\pi G\, r}  \left(1 + \frac{4\, a_0 \, r^2}{G\,
M}\right)^{-1/2}.\end{equation}
Defining
\begin{equation}
b  := \sqrt{\frac{G\, M}{4\, a_0}} 
\end{equation}
this becomes
\begin{equation} \label{pp-dmB}
\rho_{\rm d} = \frac{\rho_0}{s\, \sqrt{1+s^2}}
\end{equation}
where $s = r/b$ and
\begin{equation}
\rho_0 = \frac 1 \pi \left(\frac{a_0}{G\, M}\right)^{\frac 3 2} \, M = \frac{1}{8\,\pi} \cdot
\frac{M}{b^3}.
\end{equation}
One should note the scalings:
\begin{align}
b & \propto M^{\frac 1 2}, \\
\rho_0 & \propto M^{-\frac 1 2}, \\
\rho_0\, b^3 & \propto M.
\end{align}
For $a_0 = 1.0\times 10^{-10}\, {\rm m}\, {\rm s}^{-2}$ and the solar mass 
$M = M_\sun = 2.0 \times 10^{30}\, {\rm kg}$,
we get
\begin{align} 
 b & = 5.8 \times 10^{14}\, {\rm m} = 1.9 \times 10^{-2}\, {\rm pc},
 \\
\rho_0 & = 4.1 \times 10^{-16}\, {\rm kg}\, {\rm m}^{-3} = 6.0 \times 10^3\, M_\sun\, {\rm pc}^{-3}.
\end{align}
\par We see  that for a point mass,  the corresponding dark matter density 
$\rho_{\rm d}$ given by (\ref{pp-dmA}) or (\ref{pp-dmB}) behaves as 
$r^{-2}$ for large $r$, so that 
\begin{math} \lim_{r\to\infty} r^2\, \rho_{\rm d}(r) = \rho_0 \end{math}, and 
 the mass content
$\int \rho_{\rm d}\, d^3r$ diverges linearly.  Unlike conventional dark matter theories which could 
circumvent this infinity by stating that $\lim_{r\to\infty} r^2\, \rho_{\rm d}(r) = 0$, in MI, 
the behavior of $\rho_{\rm d}$ for large $r$ is dictated by $\nu(h)$, which is uniquely determined by
the function $\mu(a)$.
As far as the asymptotic behavior of $\nu(h)$ for small values of $h$ is $\nu(h) \propto h^{-1/2}$ we get
$\rho_{\rm d}\propto r^{-2}$ for large $r$.  Both simple and standard forms of $\mu(a)$ lead to this asymptotic
behavior (\ref{asym1}).  In fact, if $\mu(a)\propto a$ for $a\ll a_0$,
as was proposed explicitly by \citet{Milgrom-1983ApJ},
then it is easy to see that $\nu(h)\propto h^{-1/2}$ for $h\ll a_0$.
Therefore, there is no escape from this linear divergence of mass content in MI.

\subsection{Non-conservation of momentum}
As was pointed out by \citet{Felten-1984}, momentum of a two body system is not conserved in MI.
Using the equivalent system $\vec a = \vec g$, where $\vec g = \vec h\, \nu(h)$, we can easily see why this
is so.  Consider two bodies, with masses $m_1$ and $m_2$, a distance $r$ apart.  In MI one assumes that
at the position of $m_2$ we have $\vec h_2 = -G\, m_1/r^2\, \Hat{n}$, and at the position of $m_1$ we have
$\vec h_1 = +G\,m_2/r^2\, \Hat{n}$, where $\Hat{n}$ the the unit vector joining 1 to 2.  
The forces therefore are
\begin{align} \vec F_{1\to 2} & = -\frac{G\, m_1\, m_2}{r^2}\, \nu_1  \, \Hat{n},
\\ \vec F_{2\to 1} & = +\frac{G\, m_2\, m_1}{r^2} \, \nu_2  \, \Hat{n}.
\end{align}
where
\begin{math} \nu_1  = \nu(G\, m_1/(r^2\, a_0)) \end{math},
\begin{math} \nu_2  = \nu(G\, m_2/(r^2\, a_0)) \end{math}.
Since $\nu_1\neq \nu_2$, Newton's third law is violated, and momentum is not conserved.
\subsection{Non-conservation of energy}
If $\rho_{\rm m}(\vec{r})$ has spherical symmetry,  $\vec{h}$ is
 parallel to $\Hat{r}$, and $\vec{h} \times
\vec\nabla \nu(h)$ vanishes.  But in general, we do not have
spherical symmetry.
Consider for example the gravitational field of a spheroid.
Relative to the center, the first two terms of the multipole expansion of the potential 
consists of a monopole of  mass $M$ and a quadrupole of moment $\epsilon\, M\, d^2$, 
where $d$ is the length scale of the quadrupole moment,
and $\epsilon=\pm 1$ for prolate or oblate spheroids. 
In spherical polar coordinates $(r, \theta, \phi)$,
we can write the Newtonian potential for large $r$ (see appendix \ref{appendix2})
from which we find:
\begin{equation}\label{eq:curlg}
 \vec\nabla\times\vec g =  -\frac 3 4\,\epsilon \sqrt{G\, M\, a_0} \cdot\frac{d^2}{r^4}\, 
\sin(2\,\theta) \, \Hat{e}_{\varphi} + \Order{1}{r^6},
\end{equation} 
valid for  $r \gg 2\, b = \sqrt{G\, M/a_0}$ (see \ref{eq:rc}).
A non-vanishing curl of the $\vec g$ field would imply some
effects, because from $\vec{a} = \vec{g}$, one could easily
get the work-kinetic energy theorem 
\begin{equation}
  \Delta \left( \tfrac 1 2 \, m \, v^2 \right) = m \, \int \vec{g} \cdot d\vec{r}. 
\end{equation}
If $\vec\nabla\times\vec{g} \neq 0$, then the line integral
depends on path, and in particular it does not vanish for a closed
path.   Let's define $\delta\vec g$ by
\begin{equation}  \vec g = \vec h + \delta\vec{g}, \end{equation}
since $\vec\nabla\times\vec h = 0$ we have
\begin{equation}
\oint \vec\nabla\times\vec g \cdot d\vec{r} = \oint \vec\nabla\times\delta\vec{g} \cdot d\vec{r}.
\end{equation}
Therefore, the work-kinetic energy theorem could be written as
\begin{equation}
 \Delta \left( \tfrac 1 2 \, m \, v^2 + V \right) =
m \, \int \delta\vec{g} \cdot d\vec{r}, 
\end{equation} where $V$ is the
usual Newtonian potential energy defined by 
$m\,\vec h = - \vec\nabla V$. If
$\delta\vec{g}$ is small, we can consider the right hand side as
a perturbation and deduce some observable results. 
\subsection{Effect on a binary system}
Before going to the subject, it should be noted that the investigation by \citet{ZLB-2010} on 
modified Kepler's law and two-body problem in MOND, is in the context of a conservative theory given by
a Lagrangian; however, the treatment we present here is in the context of MI.
\par In MI theory, 
consider two objects with masses $m_1$ and $m_2$ forming a binary, and in free fall
 in the gravitational field $\vec h$
of a galaxy. Transforming to the equivalent system (\ref{eq:demond1}-\ref{eq:demond3}),
and \emph{assuming that for the internal dynamics of the binary we can use ordinary Newtonian 
gravitation},  we get the following differential
equations:
\begin{align}
m_1\, \ddot{\vec r}_1 & =+G\, m_1\, m_2\, \frac{\vec{r}}{r^3} + m_1\, \vec g(\vec r_1),
\\
m_2\, \ddot{\vec r}_2 & =-G\, m_1\, m_2\, \frac{\vec{r}}{r^3} + m_2\, \vec g(\vec r_2),
\end{align}
where $\vec r = \vec r_2 - \vec r_1$.  It follows from these two equations that
\begin{equation}
\ddot{\vec{r}} = -G\, (m_1 + m_2) \frac{\vec{r}}{r^3} + \vec a_{\text{T}},
\end{equation}
where $\vec{a}_{\text{T}}$ is the galactic \emph{tidal} field 
\begin{align}
\vec{a}_{\text{T}} & = \vec{r}\cdot\vec{\nabla} \vec{g}(\vec{R}),
\end{align}
$\vec R$ being the center of mass of the binary.

In the Newtonian theory, $\vec\nabla\times \vec{g} = 0$, and it follows that 
$\vec\nabla\times\vec{a}_{\text{T}}$ also vanishes, because
\begin{equation} \vec\nabla \times \vec a_{\text{T}}\Big\vert_{\vec r} 
= \vec\nabla\times\vec g\Big\vert_{\vec R}. 
\end{equation}
In MI however,
because of (\ref{eq:demond3}), the galactic tidal field is not conservative.

If $\Hat{n}$ is the unit vector normal to the orbital plane, along the angular momentum of the binary, then
the work done on the binary by the tidal field of the galaxy, in one revolution would be
\begin{align}
\Delta W  = \reducedmass \oint \vec a_{\text T}\cdot d\vec r 
& = \reducedmass \iint \vec\nabla\times\vec a_{\text{T}}
\cdot \Hat{n} \, da
\\ & = \reducedmass \iint \vec\nabla\times\vec g
\cdot \Hat{n} \, da,
\end{align}
where $\reducedmass$ is the reduced mass of the binary. 
Consider a binary with elliptical orbit---semi-major axis $s$ and eccentricity $\varepsilon$---and the 
center of mass at  $(r_0, \theta_0, \phi_0)$;
and let's approximate the gravitational field of the galaxy by a mass + quadrupole $(M, \epsilon\, M\, d^2)$
at the origin. Let the angular momentum of the binary (which defines its orbital plane) makes an angle
$\beta$ with $\Hat{e}_\phi(\phi_0)$.  Using (\ref{eq:curlg}), and noting that for $s \ll r_0$ both
$\vec\nabla\times\vec g$ and $\Hat{e}_\phi$ are almost constant over the orbit of the binary, we get
\begin{align}
\Delta W & = -\frac{3\,\epsilon}{4}\, \frac{m_1\, m_2}{m_1 + m_2}
\, \sqrt{G\, M\, a_0} \,\frac{d^2}{r_0^4}\, 
\left[\pi\, s^2\, \sqrt{1-\varepsilon^2}\right]
\cr & \qquad \qquad \times \sin(2\, \theta_0)\, \cos\beta.
\end{align}
By Kepler's law, the frequency of the orbit is
\begin{equation}
f = \sqrt{\frac{G\,(m_1 + m_2)}{4\, \pi\, s^3}}.
\end{equation}
Therefore, due to the tidal force being non-conservative in MI, 
the binary looses or gains energy with power
\begin{align}
P_{\text{MI}} 
& = -\frac{3\,\sqrt\pi\,\epsilon}{8}\,  \sqrt{1-\varepsilon^2}\, \left(G\, m_1\, m_2\right)
\sqrt{\frac{M}{m_1 + m_2}}   \, \frac{d^2\, s^{1/2}}{r_0^4}
\cr & \qquad\qquad\qquad \times \sqrt{a_0}\, \sin(2\,\theta_0)\, \cos\beta.
\end{align}
This power has some peculiar properties:
\begin{enumerate}[(a)]
\item It is proportional to $\cos\beta$, so that for $\beta > 90^\circ$ the sign changes.  In other words,
the binary either looses energy or gains energy according to its sense of rotation around  
$\Hat{e}_\phi$!
\item It is proportional to $\sin(2\,\theta_0)$, which means that there is a  sign change at
$\theta_0 = 90^\circ$!
\end{enumerate}
$P_{\text{MI}}$ is proportional to $s^{1/2}$, which makes sense---it is a tidal effect.
In comparison, the power of the gravitational wave radiation of the binary as found by
 \citet{Peters-Mathews-1963} 
\citep[see also][]{Landau-Lifshitz-fields}  is proportional to $s^{-5}$ thus:
\begin{equation}
P_{\text{GW}}  = -\frac{32}{5}\cdot\frac{G^4\, m_1^2\, m_2^2\, (m_1 + m_2)}{c^5\, s^5} 
\cdot \frac{1 + \frac{73}{24} \, \varepsilon^2 + \frac{37}{96}\,\varepsilon^4}{\left(1-\varepsilon^2\right)^{7/2}}.
\end{equation}
As an example to see that this effect could be in principle observable, 
consider a binary with
$m_1   \simeq m_2$ and
$\varepsilon  \simeq 0$. For this binary we have 
\begin{equation} 
P_{\text{GW}} \simeq -\left(\frac{m}{M_\odot}\right)^5 \left(\frac{s}{{\rm au}}\right)^{-5}
 \left( 4 \times 10^{13}\right)\,{\rm W}
. \end{equation}
Now, suppose this binary is at a distance $r_0 = 15\, {\rm kpc}$ from the center of the Milky Way.
Assuming   $M = 5 \times 10^{10}\, M_\sun$,
$d = 10\,{\rm kpc}$, and $\epsilon = -1$, (see \ref{eq:M-MW}-\ref{eq:d-MW}) 
 we have $b \simeq 4\,{\rm kpc}$, and condition $r_0 \gg b$ is almost fulfilled, and we get
\begin{align}
P_{\text{MI}} &\simeq  
\left[\sin(2\,\theta_0)\, \cos\beta\right]
\left(\frac{m}{M_\odot}\right)^{\frac 3 2}\,
\left(\frac{s}{1\, {\rm au}}\right)^{\frac 1 2} 
 \,\left(\frac{a_0}{10^{-10}\, {\rm m}\, {\rm s}^{-2}} \right)^{\frac 1 2} 
 \cr & \qquad  \times \left( 2\times 10^{14}\right)\, {\rm W}.
\end{align}

The non-conservation of energy in a binary system we just presented is a consequence of
$\vec\nabla\times\vec a_{\text{T}} \neq 0$ which is a consequence of $\vec\nabla\times\vec g \neq 0$.
For an $N$-body system like a globular cluster, this means violation of the virial theorem,
which is valid in modified gravity theories \citep[see][]{Zhao2010}.

\section{Modified Inertia vs Modified Gravity}
The Poisson equation of the Newtonian gravity,
\begin{math} \nabla^2\phi = 4\, \pi\, G\, \rho \end{math}, is
obtained from the Lagrangian density
\begin{equation} \LagDensity_{\text{N}} = -\rho\, \phi - \frac{1}{8\,\pi\, G} \abs{\vec\nabla\phi}^2.
\end{equation}
\citet{BM-1984} introduced the  Lagrangian density
\begin{equation} \label{eq:BM-lagrangian}
\LagDensity_{\text{BM}} = -\rho\, \psi - \frac{1}{8\,\pi\, G} \BMF\left(\frac{\abs{\vec\nabla\psi}^2}{a_0^2}
\right),
\end{equation}
where the functions $\BMF$ and $\mu$ are related by
\begin{equation} \mu(h) = \BMF'\left({h^2}/{a_0^2}\right), \quad \BMF'(x) = \oder{\BMF}{x}, \end{equation}
and $\psi$ now satisfies the following equation:
\begin{equation}
\vec\nabla\cdot \left[ \mu\left(\abs{\vec\nabla\psi}/a_0\right) \, \vec\nabla\psi\right] = 4\, \pi\, G\,
\rho.
\end{equation}
For $\abs{\vec\nabla\psi} \gg a_0$, we have $\mu \simeq 1$, and we get the usual Poisson equation for
the Newtonian gravitational field; but for $\abs{\vec\nabla\psi} \lesssim a_0$, we have deviations from 
Newtonian gravity, consistent with rotation curves of spiral galaxies.  This theory is called the
modified gravity version of MOND.


Let's fix our terminology and notation.  We have three models:
\begin{itemize}
\item[MI] The modified inertia theory of \citet{Milgrom-1983ApJ}, summarized by equations 
(\ref{eq:mond1}-\ref{eq:mond3}).
\item[BM] The modified gravity of \citet{BM-1984}, given by 
$\LagDensity_{\text{BM}}$ (eq~\ref{eq:BM-lagrangian}).
\item[NN] The Non-conservative Newtonian dynamics given by equations 
(\ref{eq:mond2},\ref{eq:mond3}, \ref{eq:demond1}-\ref{eq:demond3}).
\end{itemize}

In section~\ref{sec:dual} we showed that MI is equivalent to NN.  But BM could not be equivalent to NN,
because if we write the usual Lagrangian for the motion of a test particle in BM model as
\begin{equation} \label{eq:particle-lagrangian}
 L = \frac 1 2\, m\, v^2 - m\, \psi, \end{equation}
we get the equation of motion as
\begin{equation} \vec a = -\vec\nabla\psi. \end{equation}
But the equation of motion of the NN model is $\vec a = \vec g$, and we know that $\vec g \neq -\vec\nabla\psi$,
simply because $\vec\nabla\times\vec\nabla\psi = 0$, but $\vec\nabla\times\vec g \neq 0$.

As far as $\psi$ is independent of time, 
from (\ref{eq:particle-lagrangian}) it follows that in BM energy of a test particle is conserved.
On the other hand, MI theory is equivalent to NN for which 
$\vec\nabla\times\vec g \neq 0$ (\ref{eq:demond3}), which means
the potential is not well defined and the energy is not conserved.

\section{Summary and Conclusion}
Modified Inertia  of Milgrom states that if $\rho_{\rm m}$ is the mass distribution of a galaxy,
the acceleration $\vec a$ of a test particle (a star) is the solution of
\begin{math} \vec a\, \mu(a) = \vec h\end{math}, where $\mu(a)$ is a function characterizing the
theory (see appendix \ref{appendix1}) and $\vec h$ is the gravitational field satisfying
\begin{math}   \vec\nabla\cdot\vec{h}  = - 4 \, \pi \, G \, \rho_{\rm m} \end{math},
and \begin{math}   \vec\nabla\times\vec{h}  = 0\end{math}.  Introducing the function $\nu(h)$ 
(see appendix~\ref{appendix1}) we have $ a = h\, \nu(h)$.
Introducing $\vec g = \vec h\,\nu(h)$, the dynamics of the particle is given by 
the Newtonian dynamics equation $\vec a = \vec g$, where $\vec g$ satisfies 
\begin{align} \tag{\ref{eq:demond2}}
\vec\nabla\cdot\vec{g}  &=  - 4 \, \pi \, G \, \rho_{\rm m} \, \nu(h) +
\vec{h} \cdot \vec\nabla\nu(h),
\\ \tag{\ref{eq:demond3}}
\vec\nabla\times\vec{g} & =  - \vec{h} \times \vec\nabla \nu(h).
\end{align}
This means that modified inertia version of MOND is mathematically equivalent to an acceleration  field with three 
features: 
\begin{enumerate}[(1)]
  \item  the mass density is modified, \begin{math} \rho_{\rm m} \to \rho_{\rm m}\, \nu(h) \end{math}, 
  \item  there is a dark matter with density
         \begin{math}\rho_{\rm d} = -\nu'(h) \, \vec h \cdot \vec\nabla h/(4\,\pi\, G) \end{math},
  \item the acceleration field is non-conservative.  This leads to non-conservativeness of the galactic tidal force
  which has some observable effects on the binaries and perhaps globular clusters of stars.
\end{enumerate}
Besides, it was argued that the modified gravity theory of \citet{BM-1984} is not exactly equivalent to
modified inertia theory of \citet{Milgrom-1983ApJ}.
It should be noted that the non-conservative Newtonian system given in this article
is different from the quasi linear MOND of \citet{Milgrom-2010a}, which is conservative.

Modifications introduced by Milgrom started a fruitful investigation by 
various researchers which still continues, and this is invaluable.   What we are saying in this article, 
is that some models could be interpreted in the framework (or paradigm) of Newtonian 
dynamics, which could lead to new insights.

\paragraph*{Challenges to form a conservative MI theory}
From the first days of introducing MOND by Milgrom, there has been efforts to form 
a conservative MI theory.  One way is to replace the standard action $\frac 1 2 \int v^2\, {\rm d}t$ 
by a more complicated action of the form $A_m\, S[\vec r(t), a_0]$, where $A_m$ depends on the body, related to 
particle's mass, and $S$ is a functional of the trajectory of the particle, but depending on the particle's
entire history \citep{Milgrom-1994a, Milgrom-1999}. 
Milgrom demonstrated that, in the context of such theories, the simple MOND relation (\ref{MOND}) 
is exact for circular orbits in an axisymmetric potential (although not for general orbits)
\citep{SM2002}. Such theories are usually highly non-local \citep{Milgrom-2001a, SM2002}.
Recently, \citet{Alzain:2017} has tried to construct a relativistic theory implementing Milgrom's MI.

\begin{appendix}
\section{The Interpolating Functions}  \label{appendix1}
Let's review the  transformation $\mu(a)\to\nu(h)$ to invert the interpolating function
\citep[][]{SJ2007, McGaugh-2008, Zhao2010, FM2012a}.

\begin{figure} 
\includegraphics[width = 7.2cm]{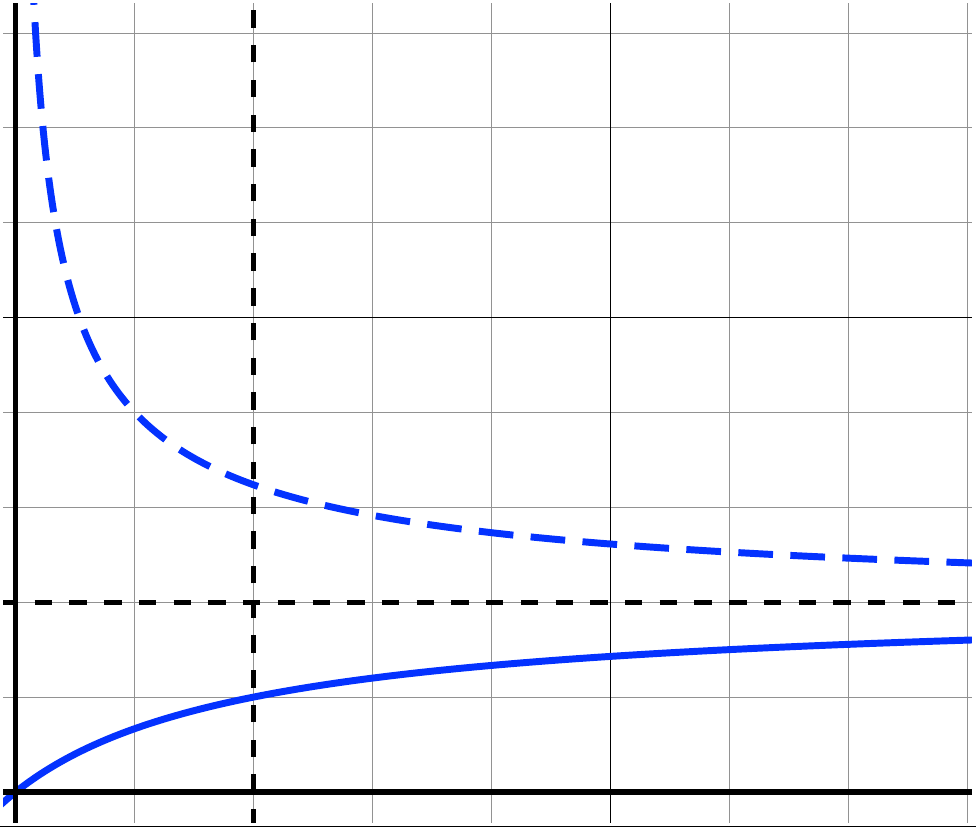}
\setlength{\unitlength}{1mm}
\put(-10,10){$\mu(a)$}
\put(-10,26){$\nu(h)$}
\caption{Simple form of the interpolating functions $\mu(a)$ (\ref{simple-mu}), solid; 
and $\nu(h)$ (\ref{simple-eta}), dashed. 
The scale $a_0$ is indicated by the vertical dashed line. The horizontal dashed line
indicates 1 on the ordinate.}
\label{fig1}
\end{figure}

The equation
\begin{math}  h = a\, \mu(a) \end{math},
where $\mu(a)$ satisfies (\ref{eq:mu-function1}-\ref{eq:mu-function3}),
 could be solved for $a$.  The proof is a simple application of the inverse function theorem.
Denoting the solution of $h = a\, \mu(a)$ by $a = f(h)$ let's define $\nu(h) = h^{-1}\, f(h)$, so
that we have:
\begin{equation} \label{eta}
a = h\, \nu(h).
\end{equation}
Dividing $a \, \mu(a) = h$ by $a = h \, \nu(h)$, one gets 
\begin{math} \mu(a)\, \nu(h) = 1\end{math},
 from which it follows that $\nu(h)$ is a dimension-less,
monotonically decreasing function, asymptotic to 1
for large $h$.
\begin{align} \lim_{h\to 0^+} \nu(h) & = +\infty, 
\\ \lim_{h\to\infty}\nu(h) &  = 1, 
\\  \nu'(h) & < 0 \quad \forall h > 0. 
\end{align}
 The \textit{simple} form of the interpolating function $\mu(a)$ is (Fig~\ref{fig1})
\begin{equation} \label{simple-mu} \mu(a) = \frac{a}{a + a_0}. \end{equation}
Solving 
\begin{math} a \, \mu(a) = \frac{a^2}{a + a_0} = h\end{math} for $a$ one gets
\begin{equation} \label{simple-eta}
\nu(h) := \frac 1 2 \, \left( 1 + \sqrt{1 + \frac{4\, a_0}{h}} \, \right).
\end{equation}
For later use, we note that the asymptotic form of $\nu(h)$ is
\begin{align}
\nu(h) &\simeq 1 + \frac{a_0}{h} \qquad h \gg a_0, \\
\label{asym1}
\nu(h) &\simeq  \sqrt{\frac{a_0}{h}} \qquad h \ll a_0,
\\ \label{eq:etaprime-simple}
\nu'(h)  & = -\frac{a_0}{h\,\sqrt{h^2 + 4\, a_0\, h}}
 \simeq -\frac{a_0^{1/2}}{2\, h^{3/2}} \qquad h \ll a_0.
\end{align}
 The \textit{standard} form of the interpolating functions $\mu(a)$ and $\nu(h)$ is the pair
\begin{equation} \mu(a)  = \frac{a}{\sqrt{a^2 + a_0^2}}, 
\quad \nu(h)  = \frac{1}{\sqrt{2}}
\left( 1 + \sqrt{1 + \frac{4 \, a_0^2}{h^2}}\, \right)^{1/2}.
\end{equation}
It should be  noted that both simple and standard forms of $\mu(a)$ lead to the same 
asymptotic behavior for $\nu(h)$ for small $h$, and therefore the same asymptotic
form for $\nu'(h)$, for small $h$.

\section{A Mass + Quadrupole System} \label{appendix2}
In spherical-polar coordinates $(r, \theta,\varphi)$,
consider the potential
\begin{equation} \psi(r, \theta) = - \frac{G\, M}{r} -\epsilon \frac{G\, M\, d^2}{r^3} 
\left(\frac 3 2 \, \cos^2\theta - \frac 1 2\right),
\end{equation} 
where $\epsilon = \pm 1$.
This potential describes a mass + a quadrupole system. 
 It is straightforward to find $\vec h = -\vec\nabla\psi$,
and then using $\vec g = \vec h\, \nu(h)$ to find $\vec g$.

\begin{align} 
\vec h  & = h_r\, \Hat{e}_r + h_\theta\, \Hat{e}_\theta \\
-\frac{h_r}{G\, M} & =  \frac{1}{r^2} + \frac{\epsilon\, d^2}{r^4}\cdot\tfrac{3}{2} (3\, \cos^2\theta - 1) \\
-\frac{h_\theta}{G\, M} &  = \frac{\epsilon\, d^2}{r^4} \frac 3 2\, \sin(2\,\theta) , \\
\frac{h^2}{(G\, M)^2} & = \frac{1}{r^4} + \epsilon\, \frac{d^2}{r^6} (3\, \cos^2\theta - 1) 
+ \Order{1}{r^8}
\end{align}
Using either the simple or the standard form of the interpolating function,
we get
\begin{equation}\label{eq:rc}
\frac{\nu'(h)}{h} \simeq \frac 1 2 \, \sqrt{\frac{a_0}{(G\,M)^5}} \, r^5, 
\qquad r \gg  \sqrt{\frac{G\, M}{a_0}} = 2\, b.
\end{equation}
Now, using
\begin{align}
\frac{h_r}{r} & = - \frac{G\, M}{r^3} + \Order{1}{r^5}, \\
\pder{h^2}{\theta} & = \epsilon\, \frac{G^2\, M^2\, d^2}{r^6} \, 3\,\sin(2\,\theta) + \Order{1}{r^8}, \\
\pder{h^2}{r} & = - \frac{G^2\, M^2}{r^5} +\Order{1}{r^7}, 
\end{align}
 we can find the leading term of $\vec\nabla\times\vec g$.
\begin{align}
\vec\nabla\times\vec g & = -\vec h \times \vec\nabla\nu(h) \\
& = -\frac{\nu'(h)}{2\, h} \vec h \times \vec\nabla h^2 \\
& = -\frac{\nu'(h)}{2\, h} \left(\frac{h_r}{r}\pder{h^2}{\theta} - h_\theta\, \pder{h^2}{r} \right) 
\Hat{e}_{\varphi}
\\ & = -\frac{3}{4}\epsilon \sqrt{G\, M\, a_0} \cdot\frac{d^2}{r^4} \,\sin(2\,\theta) \,\Hat{e}_{\varphi}
+ \Order{1}{r^6}.
\end{align}

Using the values given by \citet{Sofue-2017},
 we consider the following mass distribution for the Milky Way (excluding the halo): 
\begin{enumerate}[(i)]
\item A central black hole of mass $M_{\text{bh}} = 3.6 \times 10^{6}\, M_\sun$.
\item A spherical bulge of mass $M_{\text{b}} = 9.2 \times 10^{9}\, M_\sun$.
\item A disk of mass $M_{\text{D}} = 4.0 \times 10^{10}\, M_\sun$ 
with exponential density of length scale $a_{\text{D}} = 5.0\,{\rm kpc}$.
\end{enumerate}
From these figures we get 
\begin{align} \label{eq:M-MW}
M & = 4.9 \times 10^{10}\, M_\sun,
\\ \label{eq:I-MW}   \epsilon\, M\, d^2 & = -5\, M_{\text{D}}\, a_{\text{D}}^2 = 
-5.0 \times 10^{12}\, M_\sun\, {\rm kpc}^2,
\\  \label{eq:d-MW}
d & =  10\,{\rm kpc}.
\end{align}

\end{appendix}


\paragraph{acknowledgements}
We would like to thank A.-H. Fatollahi for his valuable comments, which motivated the early version 
of this article.   
This work was partially supported by Alzahra University's
research council, and partly by Nazarbayev University.

%

\end{document}